\documentclass{emulateaph1}
\usepackage{graphicx}
\begin{document}
\bibliographystyle{apj}

\title{Prominence Mass Supply and the Cavity}
\shorttitle{Prominence Cavity Connection}

\author{Donald J. Schmit$^{1}$, S. Gibson$^2$, M. Luna$^3$, J. Karpen$^4$, D. Innes$^1$}
\altaffiltext{1}{Max Planck Institute for Solar System Research}
\altaffiltext{2}{High Altitude Observatory, National Center for Atmospheric Research}
\altaffiltext{3}{Instituto de Astrofisica de Canarias}
\altaffiltext{4}{NASA Goddard Space Flight Center}
\shortauthors{Schmit et al.}

\begin{abstract}
A prevalent but untested paradigm is often used to describe the prominence-cavity system: the cavity is under-dense because it is evacuated by supplying mass to the condensed prominence.
The thermal non-equilibrium (TNE) model of prominence formation offers a theoretical framework to predict the thermodynamic evolution of the prominence and the surrounding corona.
We examine the evidence for a prominence-cavity connection by comparing the TNE model with diagnostics of dynamic extreme ultraviolet emission (EUV) surrounding the prominence, specifically prominence horns.
Horns are correlated extensions of prominence plasma and coronal plasma which appear to connect the prominence and cavity.
The TNE model predicts that large-scale brightenings will occur in the SDO/AIA 171\AA~bandpass near the prominence that are associated with the cooling phase of condensation formation.
In our simulations, variations in the magnitude of footpoint heating lead to variations in the duration, spatial scale, and temporal offset between emission enhancements in the other EUV bandpasses.
While these predictions match well a subset of the horn observations, the range of variations in the observed structures is not captured by the model.
We discuss the implications of our one-dimensional loop simulations for the three-dimensional time-averaged equilibrium in the prominence and the cavity.
Evidence suggests that horns are likely caused by condensing prominence plasma, but the larger question of whether this process produces a density-depleted cavity requires a more tightly constrained model of heating and better knowledge of the associated magnetic structure.
\end{abstract}

\section{Introduction}
Prominences are high-density, low-temperature structures that are suspended in the corona.
They are often surrounded by a low-density coronal structure known as a cavity \citep{saito_68}.
Two fundamental unsolved problems in prominence physics are: how is the prominence supported, and how is mass supplied to the prominence?
The commonly accepted answer to the first question is that the magnetic field supports the prominence mass, and must be dipped to collect stable, condensed plasma.
Two models were proposed for the geometry of dipped fields: sheared arcades \citep{kippenhahn, antio_94} or flux ropes \citep{kupraad,lowhund}.
The flux rope model may also explain the morphology of the cavity: helical field lines on the interior of the flux rope form an elliptical structure when viewed down the axis of the flux rope \citep{gibfan_06b}. 
The 3D magnetic geometry of the prominence-cavity system has been addressed through MHD \citep{fangib_03, devore_00} and magnetofrictional models \citep{mackayvanball_01}, neither of which has incorporated an accurate coronal energy treatment thus far.\\\indent
The question of how mass is supplied to the prominence has also been theoretically addressed. 
The thermal instability of coronal energy balance suggests that coronal equilibrium is susceptible to large-wavelength, isobaric condensing perturbations \citep{field_65}. At present, the thermal non-equilibrium (TNE) model represents the most detailed theoretical explanation of the formation of a prominence through a form of thermal instability.
This model predicts that plasma condensations form through time-dependent evaporation and cooling driven by heating localized at loop footpoints \citep{mok_90, antio_91,kuin_82}.
The geometry of each flux tube dictates the properties of each condensation, including whether condensations form and remain stationary or undergo a cycle of formation and destruction by falling onto the chromosphere \citep{karpen_01, karpen_05}, while the heating timescale determines whether condensations form at all \citep{karpen_08}. 
TNE in prominences, coronal loops, and coronal rain has been studied in detail using 1D hydrodynamic and multidimensional MHD simulations \citep{karpen_03, karpen_06, muller_03, mok_08, luna_12}. 
The dynamic evolution of density and temperature determines the emissivity of the plasma along the loop, which in turn is directly comparable to observations through forward modeling. 
\\
\begin{figure}
\center
\includegraphics[width=0.4\textwidth]{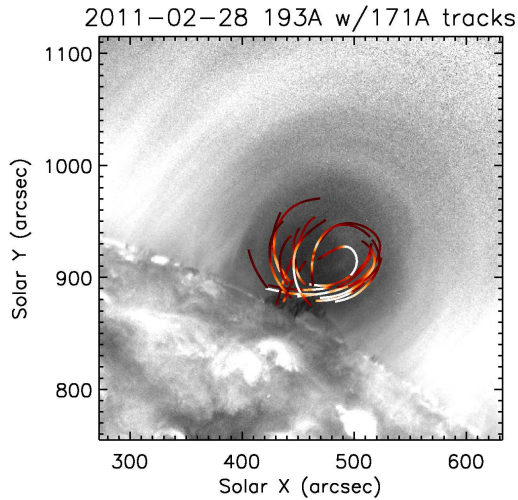}
\caption{The overlay of 171\AA~horns over a 193\AA~image of the cavity observed 2011 February 28. A detailed discussion of the extraction and analysis of horn structures is presented in Paper 1.}
\label{fig:o171}
\end{figure}
\indent Many observational studies have constrained the plasma properties within the cavity as well as the prominence-corona transition region (PCTR).
The large scale temperature and density structure of the cavity has been documented by \cite{schmit_11} and \cite{kucera_12}.
Emission inside the cavity from ion species characteristic of hot plasma (relative to quiet sun) have been observed by \cite{habbal_10} and \cite{reeves_12}.
The dynamics and fine structure of the PCTR have been discussed by \cite{heinzel_01} and \cite{heinzel_08}, in which their analysis focuses on hydrogen and helium lines and continua forming between the $10^4$ K condensation and the $10^6$ K corona.
 A complete review of spectral diagnostics on prominences and their coronal environment is found in \cite{labrosse_10}.\\\indent
In \citet{schmit_13}, hereafter referred to as Paper 1, we characterized the properties of time-dependent EUV emission in a prominence and the associated cavity observed by the Solar Dynamics Observatory Atmospheric Imaging Assembly \citep{lemen_12}.
In this research, we compare these observations with emission structures predicted by the TNE model, to derive insight into the poorly-understood relationship between the prominence and cavity.

In Section 2, we briefly summarize the observation of prominence horns. Section 3 describes the hydrodynamic simulation, the emission structures that are produced during the TNE process, and how that emission compares with horn observations.
In Section 4, we consider the roles of horns and the TNE process in producing the low-density cavity.
Our conclusions and the next steps necessary to understand the dynamic prominence-corona system are summarized in Section 5.

\section{Horn Observations}
Paper 1 details the diagnostic methods used to characterize prominence horns from multi-bandpass SDO/AIA datasets, so we summarize those results briefly here.
The dominant components of emission variability in the prominence-cavity region are prominence horns (Paper 1, Section 2).
Figure \ref{fig:o171} shows an overlay of the outlines of several horns extracted from a cavity on 2011 February 28.
Horns are most clearly visible in the 171\AA~bandpass as strong brightenings that extend from the prominence into the cavity, achieving heights similar to the 50 Mm coronal hydrostatic scale height.
The 171\AA~brightenings were correlated with changes in the perimeter of the prominence, such that 304\AA~emission would appear to extend up the base of the horn, and were accompanied by weak 193\AA~bandpass brightening.
Temporally, we found that that the peak brightness of the 171\AA~emission preceded the peak extension of the 304\AA~prominence by 30 $\pm$ 18 min, whereas a -0.6 $\pm$ 23 min minute time lag existed between 193\AA~and 171\AA~emission.\indent
\begin{figure*}
\center
\includegraphics[width=.8\textwidth]{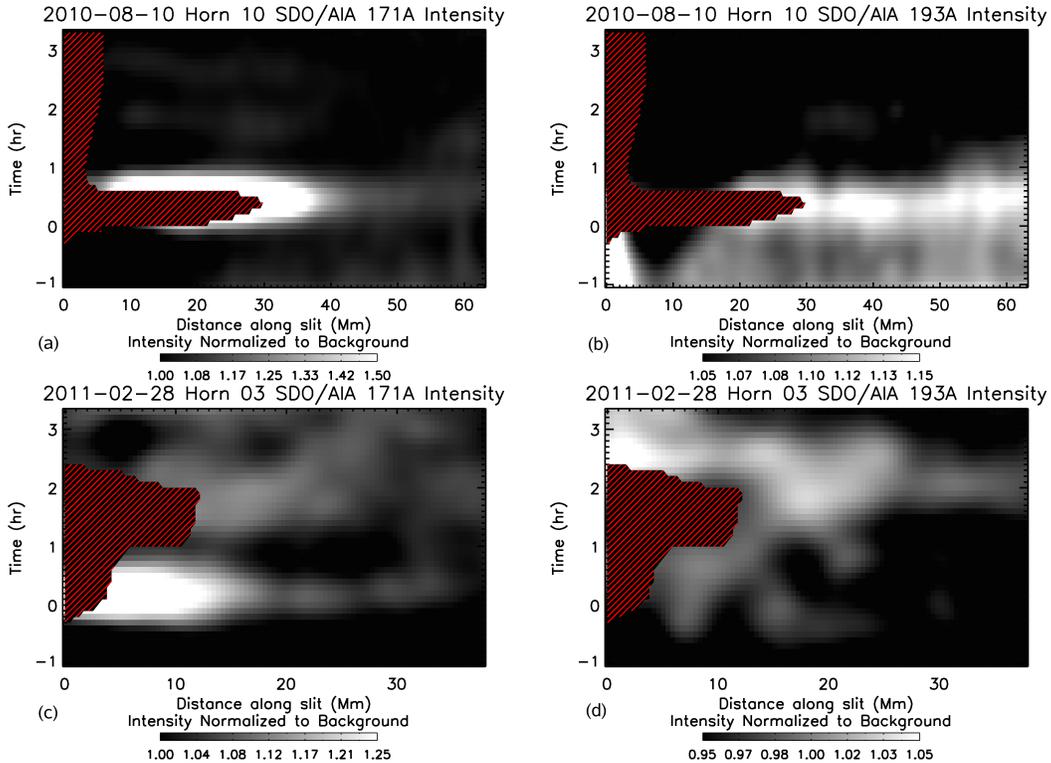}
\caption{Two extracted horn datasets presented in a time-distance plot. The x-axis is position along the slit (see Paper 1, Section 3.1), and the y-axis is time. (a) and (c) show 171\AA~emission in grayscale, (b) and (d) show 193\AA~emission in grayscale. The red-hatched region denotes where strong 304\AA~emission is observed.}
\label{fig:horn}
\end{figure*}
In addition to temporal and spectral information, prominence horns were found to have a particular geometry.
A majority of prominence horns exhibited a concave-up geometry, which appeared constant over the duration of the feature.
Because these features are likely to be low-$\beta$ field-aligned structures, they can be modeled with one-dimensional (1D) hydrodynamic simulations of plasma within a coronal flux tube. 
To present the data in a manner which is most easily comparable to our loop models, we created time-distance diagrams similar to those in \cite{karpen_05} and \cite{berger_12} where the temporal evolution of emission along a curved slit is plotted against distance along the slit.
Figure \ref{fig:horn} shows two examples of the 45 horns that were analyzed in Paper 1, illustrating their fundamental characteristics and the variations  between individual horns.
Figures 2a and 2c show the correlation between extensions of the prominence in 304\AA and larger brightenings in the 171\AA~bandpass.
The durations of the 171\AA~brightenings for these two horns are nearly identical despite their different lengths.
Both brightenings occur simultaneously across the entire horn, so there is not a clear prominence-to-cavity or cavity-to-prominence progression.
However, these horns differ significantly in the relationship between the 171\AA\ and 304\AA~emissions.
For Horn 10, the 304\AA~emission reaches maximum extension 0.2 hr before the peak emission in 171\AA, while
for Horn 3, the 304\AA~emission goes through two distinct periods of motion: $0 \le t \le 0.6$ hr and 1 hr $\le t \le 2$ hr.
The peak extension (which lasts from t=1.2hr to t=2hr) occurs 0.6 hr after the peak emission in 171\AA.
This variation in timing is reflected in the temporal error bars presented in Paper 1 (Section 3.2).\indent
Figures~\ref{fig:horn}b and \ref{fig:horn}d are time-distance diagrams for 193\AA~emission, also displaying the 304\AA~emission in red hatching.
In both horns, the peak 193\AA~emission is significantly dimmer than the 171\AA~emission, and the position of peak 193\AA~brightness sits higher along the slit than the 171\AA~feature.
In Horn 10, the peak 193\AA~emission occurs approximately 0.1 hr before the 304\AA~peak extension (171\AA~and 193\AA\ peaks are simultaneous).
In Horn 3, the peak 193\AA~emission occurs 0.2 hr after the peak 304\AA~extension, and 1.8 hr after the 171\AA~peak.
Thus, the two horns displayed in Figure~\ref{fig:horn} exhibit different correlations between the cooler 171\AA~emission and the hotter 193\AA\ emission. \indent

In summary, we have found that the properties of the 171\AA~emission in horns are roughly uniform throughout the datasets: peak intensity and duration do not vary greatly among horns despite differences in length, and the emission does not exhibit a preferred propagation direction. However, we find significant variation in the correlations among the 171\AA, 304\AA, and 193\AA~emissions.
Intense brightenings in the 171\AA~are unique to that bandpass, and the properties of that emission are largely uniform.
The other bandpasses exhibit a range of temporal associations with the 171\AA~feature but a more consistent spatial association.
As we compare these observations to models of the hydrodynamic processes, the degree of temporal correlation among these varying emissions will be a primary constraint.
\section{Thermal Non-equilibrium and Prominence Horns}
The corona is dominated by the magnetic field, and the timescales for its non-eruptive evolution ranges from days to weeks.
The thermodynamic evolution time is very rapid in comparison, as thermal conductivity is very efficient at transporting energy along the magnetic field and the coronal radiative loss time is also relatively fast.
Assuming a constant density of $10^8$ cm$^{-3}$ and a radiative loss rate of $10^{-21}$ erg s$^{-1}$ cm$^3$, it would take approximately 2000 s to cool $10^6$  K plasma to $10^5$ K.
Thus, coronal loop models that solve for the evolution of plasma contained within a fixed magnetic geometry are appropriate for studying the observed prominence and cavity plasmas. 

As in prior investigations of prominence mass formation and evolution \citep[e.g.,\ ][]{karpen_06, luna_12}, we have used the ARGOS code \citep{antio_99} and associated post-processing routines to study the dynamics of a catastrophically cooling loop and to synthesize its appearance in EUV emission lines. In contrast to earlier TNE studies, however, the present work is focused on comparing the predicted emission with multiwavelength observations of prominences, prominence horns, and the surrounding cavities, as presented in Paper 1.  
\begin{figure*}
\center
\includegraphics[width=.8\textwidth]{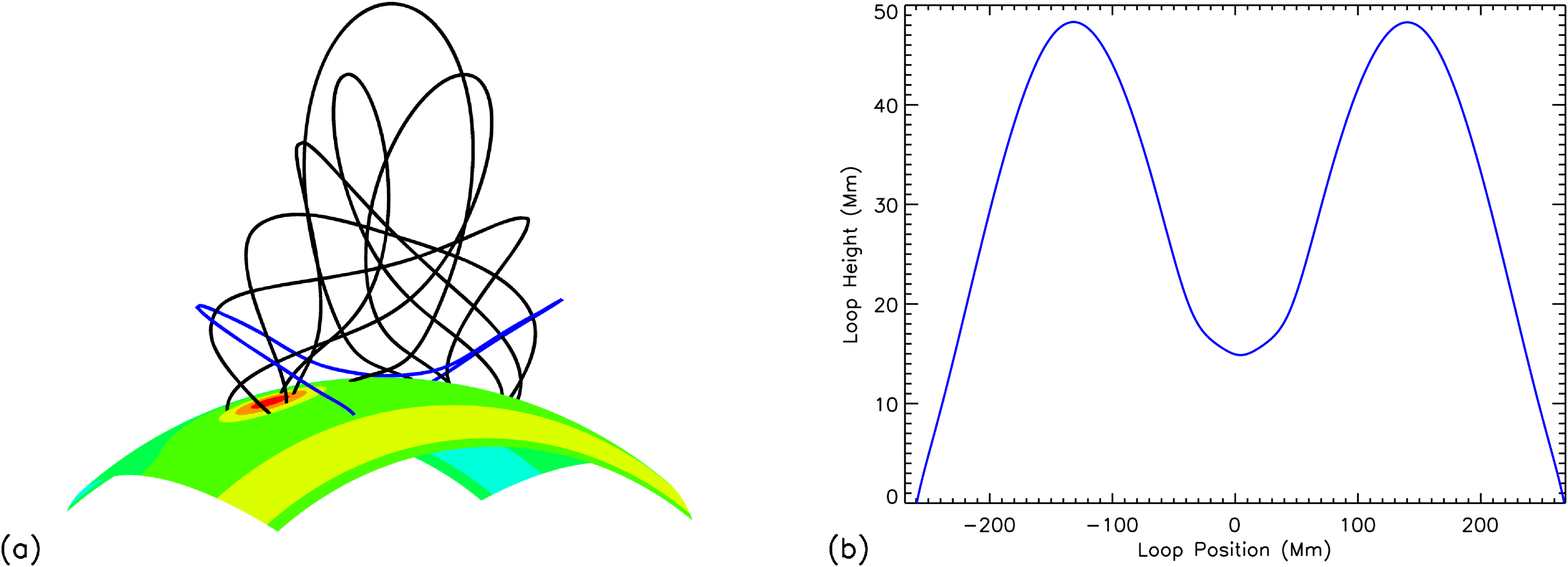}
\caption{The geometry of a 1D symmetric, dipped field line used for the TNE loop simulations. (a) Representative field lines (black) and photospheric magnetic field (color surface) from the \citet{gibfan_06b} flux rope; the extracted field field line is shown in red. (b) Height as a function of position along the extracted field line.  }
\label{fig:geo}
\end{figure*}

\subsection{Model Initial Conditions}
For purposes of this comparison, we have extracted a deeply dipped field line from a stable equilibrium of a partially emerged 3D flux rope \citep{gibfan_06b}, which is shown in Figure \ref{fig:geo}.
Condensed plasma can collect and maintain a quasi-stable position within the deep magnetic dip.
Most of the dipped field line geometries from this flux rope are asymmetric, with a high-arched side and a low-flat side.
Although those magnetic geometries are more prevalent in the 3D flux-rope model, we have used a nearly symmetric loop for simplify the analysis of the loop emission..
The ARGOS code solves the following hydrodynamic equations:
\begin{eqnarray}
\frac{\partial}{\partial t}\rho +\frac{\partial}{\partial s}\rho V=0\\
\frac{\partial }{\partial t}(\rho V)+\frac{\partial}{\partial s}(\rho V^2+P)=-\rho g_{\parallel}(s)\\
\frac{\partial}{\partial t} U+\frac{\partial}{\partial s} \left(U V-10^{-6} T^{5/2} \frac{\partial}{\partial s}T\right)\nonumber \\
=-P\frac{\partial}{\partial s}V+E(t,s)-n^2\Lambda(T)
\end{eqnarray}
where $s$ is distance along the field line, $t$ is time, $\rho$ is mass density, $V$ is the velocity, $P$ is the gas pressure, $g_\parallel$ is the component of gravity parallel to the slope of the field line, $U$ is the internal energy, $E$ is the heating function, and $\Lambda$ is the radiative loss function \citep[see][for details]{antio_99}.
The loop code uses the radiative loss function of \cite{klimcar}.
This loss function was also used in \cite{karpen_05}, and those authors found that form of radiative loss function can impact the rate of condensation.
While ARGOS is capable of including the effects of variations in the cross-sectional area of the flux tube, for simplicity in this work we have assumed the flux tube has a uniform cross-section.
Previous studies have shown that variations in cross-section may affect the rate of condensation \citep{lionello_13}.
\indent

Our dipped loop is initialized with a uniform heating rate of $6\times10^{-6}$ erg cm$^{-3}$ s$^{-1}$, and the calculation is continued for 10$^5$ s until hydrostatic equilibrium is reached. Over the next 1000 s, the heating is linearly changed to a combination of strong footpoint heating $E_1$ and weak uniform heating $E_0$ in the form
\begin{equation}
E(s)=E_0+E_1\exp[-s_t/\lambda]
\end{equation}
where $E_0$ is $6\times10^{-7}$ erg cm$^{-3}$ s$^{-1}$, $s_t$ is the position relative to the nearest footpoint, and $\lambda$ is 25 Mm.
This choice of $\lambda$ is larger than that used in previous experiments due the longer loop used for the model.
The ratio of length-to-heating scale is 21, which comparable to \citet[][ratio of 22]{karpen_01} and \citet[][ratios between 8 and 50]{muller_04}.
Very short scale lengths can lead to formation of multiple condensations \cite[see][Figure 13]{muller_04} which would complicate the emission analysis.
\indent

 By initiating with a high uniform heating, we ensure that the initial relaxed loop achieves coronal temperatures. The present study is concerned only with the plasma characteristics during and after condensation, and not with the buildup period. Therefore, when the footpoint heating is applied the background heating rate $E_0$ is lowered by an order of magnitude compared with the initial equilibrium value, in order to obtain a condensation as rapidly as possible. The lower value of $E_0$ would produce a loop center density and temperature of 2.3$\times10^7$ cm$^{-3}$ and 0.87 MK, respectively, with an integrated uniform heat input of 3$\times10^4$ erg cm$^{-2}$ s$^{-1}$ (well below typical coronal values).
\indent
\begin{figure*}
\center
\includegraphics[width=.75\textwidth]{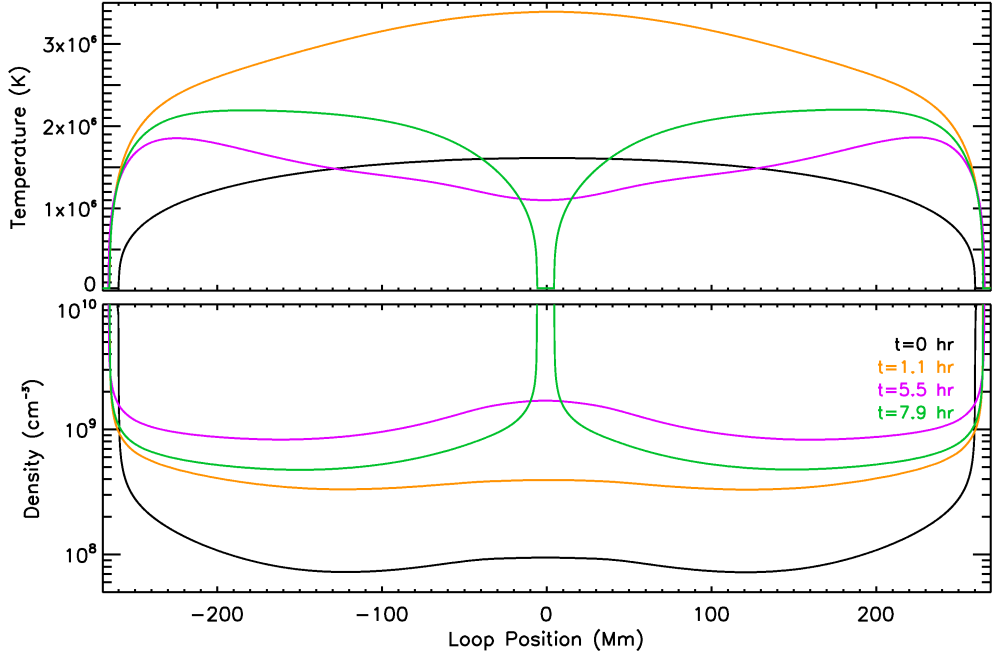}
\caption{Density and temperature along the field line at four representative times in Run A: initialization (black line), temperature maximum (orange), cooling (purple), and post-condensation equilibrium (green).}
\label{fig:evo}
\end{figure*}

In our analysis, the loop geometry and the basic form of the heating function remain unchanged so we can focus primarily on the effects of varying $E_1$.
We performed two model runs: Run A uses $E_1=1.6\times10^{-3}$erg cm$^{-3}$ s$^{-1}$ while Run B uses $E_1=2\times10^{-4}$ erg cm$^{-3}$ s$^{-1}$ . 
Runs A and B thus have energy inputs of 8$\times10^6$ and 1$\times10^6$ erg cm$^{-2}$ s$^{-1}$, respectively, comparable to standard coronal-heating estimates \citep{withbroe_77}. In both cases the footpoint-heating term dominates the integrated heat input. The variation in footpoint heating magnitude between the models results in different thermodynamic loop evolution, which affects the predicted emission structures that are compared with horn observations.
We have examined a range of $E_1$ that produces post-condensation coronal plasmas spanning the temperature range between the peak temperature responses of the AIA 171\AA~and 193\AA~bandpasses.
This allows us to map the relative response in these two bandpasses for the full range of coronal loop temperatures that could be present in the quiescent cavity (0.8 MK to 2.0 MK).
Much cooler plasmas are present during the dynamic cooling phase.

\subsection{Simulation Results}
The model coronal loop evolves through three phases, illustrated by Figure \ref{fig:evo}. 
The footpoint heating is turned on at $t=0$ (Fig. \ref{fig:evo}, black curves), causing a rapid temperature increase that is proportional to $E_1\lambda$.
Within 1-2 hours, Run A reaches a peak temperature of 3.4 MK  (orange curves) while Run B reaches 1.7 MK. 
The second phase (purple curves) is characterized by increasing density in the coronal segment of the loop, evaporated from the chromosphere, accompanied by increased radiative losses and large-scale (loop-wide) cooling. 
This cooling phase ends after $\sim$4-10 hours with a rapid influx of plasma into the loop midpoint, forming a condensation. 
During the third phase (green curves) the loop reaches a new equilibrium with hot coronal plasma on both sides of the condensation; for the dipped geometry presented in Figure \ref{fig:geo} and the symmetric heating profile of Equation 4, the condensation is gravitationally stable but continuously growing.
At t= 8hr, the loop resembles two separate coronal loops: in addition to the usual two chromosphere-corona transition regions located at the magnetic footpoints, each loop has prominence-corona transition regions located on either side of the condensation. 
\indent

In order to compare the model with our horn observations, we forward model the predicted emission from the loop simulation in each bandpass.
This forward modeling is meant to provide a rough estimate of the relative change in emission created by embedding a single dynamic loop in a quiescent corona. 
First, we convolve the density and temperature distributions with the predicted AIA response function using the Solarsoft routine ``aia\_get\_response.pro" (version 3), which specifically describes the spectral response of the instrument based on quiet-Sun level and ionization-equilibrium conditions and constant pressure.
Next, we incorporate the resulting time-dependent intensity into a 3D structural model, which allows us to consider the projection effects of the surrounding ambient corona \citep{gibson_10}.
We assume the background emission is characterized by a 53$^\circ$ long cavity at a latitude of 60$^\circ$, surrounded by a streamer with an isothermal temperature of 1.7 MK and a density profile taken from \cite{schmit_11}.
The assumed extent of the evolving loop along the line-of-sight is 6 Mm.
A study by \cite{kucera_12} found that this approach overestimates the total photometric flux of coronal structures.\indent

\begin{figure*}
\center
\includegraphics[width=.8\textwidth]{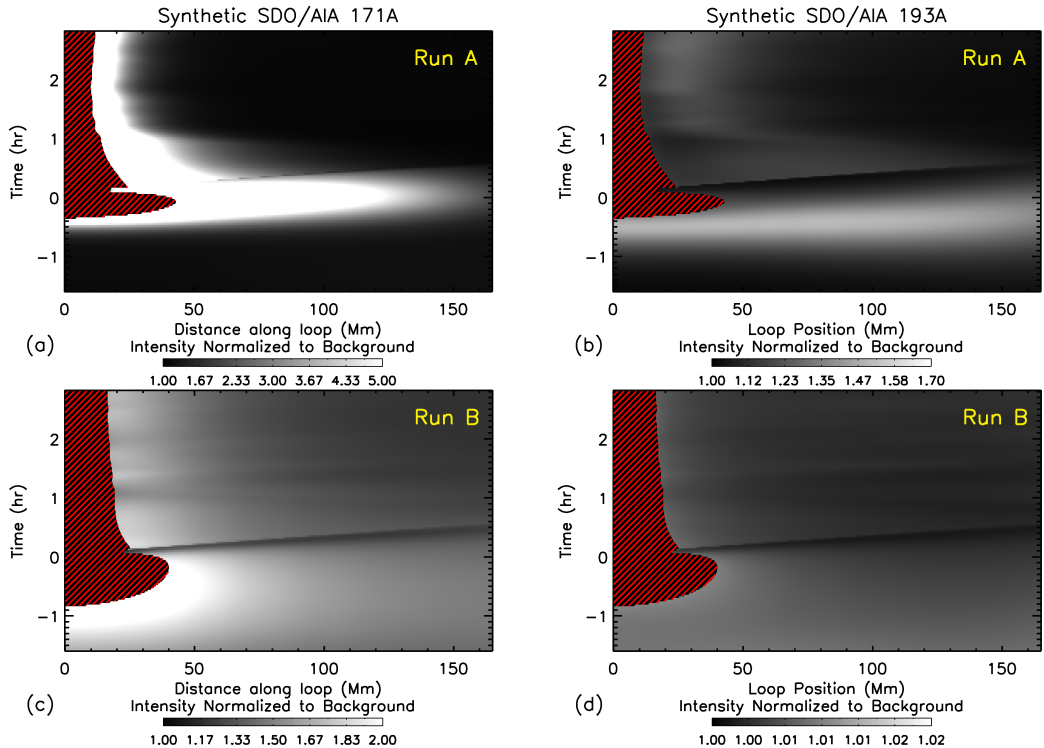}
\caption{Time-distance diagram of TNE loop simulations. As in Fig. \ref{fig:horn}, (a) and (c) show 171\AA~ emission in grayscale, while (b) and (d) show 193\AA~emission in grayscale. Regions strongly emitting in optically thin 304\AA~are shown in red-hatching. Runs A and B differ in the magnitude of $E_1$ (see Section 3.1).}
\label{fig:slit}
\end{figure*}

Figure \ref{fig:slit} shows the structures predicted by forward modeling the emission from simulation Runs A and B. We notice several similarities between the emission structures in these runs.
While both runs exhibit brightenings in 171\AA~emission and similar condensation growth (Figs. \ref{fig:slit}a and c), the intensity of the emission and the spatial extent varies; the change in 193\AA~emission (Figs. \ref{fig:slit}b and d) is significantly weaker than the change in 171\AA~emission.
Another important similarity between Runs A and B is in the temporal progression between the emission structures in different passbands: the 193\AA~emission brightens first, followed by an intensity increase in 171\AA~emission, followed by the formation of the condensation. This progression is consistent with earlier TNE studies \citep[e.g., ][]{luna_12}.
While the order of this progression is consistent, we find that the temporal offset between emissions in different bandpasses and the duration of emission strongly depends on the heating rate.\indent
Table 1 quantitatively compares model predictions of selected emission properties with their counterparts for observed prominence horns.
One of the primary differences between Run A and Run B is in  the time lag between the peak emission in 171\AA~and the formation of the condensation, $t_{171}$: stronger footpoint heating leads to a shorter time lag.
For the eight-fold decrease in localized heating $E_1$ between Runs A and B, we find that $t_{171}$ rises from 10 to 40 minutes.
These brightenings exhibit discernible rise and fall phases, which also depend on $E_1$: stronger heating yields shorter episodes of enhanced 171\AA~emission.
The duration of the brightenings are related to the broad temperature response functions for AIA and mass build up in the coronal portion of the loop.
Run A accumulates mass in the corona more rapidly than Run B and therefore cools more rapidly through the peak of the temperature response function.
$\Delta t_{171}$ measures the total duration of the brightening.
The observationally derived value of $\Delta t_{171}$ falls between the ranges provided by Runs A and B, while the observed value of $t_{171}$ is longer than our simulations.
For 193\AA, the measured and simulated quantities agree well.\indent

One of the surprising attributes of the observed horns is the relative lack of correlation between the 171\AA~and 193\AA~emission intensities, $\Delta I_{171}$ and $\Delta I_{193}$. How could a thermodynamic perturbation create a 70\% intensity perturbation in the 171\AA~bandpass and less than a 2\% perturbation in the 193\AA~bandpass, given the large overlap in temperature response of these two bandpasses? The TNE model predicts that this discrepancy can be attributed to the nature of the cooling process. 
The temperature response functions for the 171\AA~and 193\AA~bandpasses each contain a single maximum within the coronal temperature range.
The time of peak intensity for each bandpass occurs near, but not at, that maximum-emissivity temperature.
Rather, the long-duration density increase in the corona shifts $t_{171}$ and $t_{193}$ to cooler temperatures with respect to their maximum-emissivity temperatures.
The quantity $\Delta I_{193}$ is an order of magnitude smaller than $\Delta I_{171}$ because the loop mass is steadily increasing while the local temperature drops from approximately 1.6 MK (peak emissivity in 193\AA~bandpass) to 0.8 MK (peak emissivity in 171\AA~bandpass).\indent

To summarize, we have compared the synthetic emission predicted by the TNE model with the observed emission from prominence horns. 
Some model predictions match the observations quite well, demonstrating that horns are a natural consequence of the condensation process treated in our model.
In particular, the TNE model produces a 304\AA-emitting condensation embedded within a coronal loop, and a 171\AA~ brightening $\sim$50 Mm long and 3 hours in duration occurs 30 minutes before the formation of the condensation.
The model also predicts that the 171\AA~brightening is much stronger than the 193\AA~brightening, as observed.
Discrepancies also exist between the model and the observations, however, most notably in the progression of emission enhancement between bandpasses. We discuss the implications of this discrepancy in Section 5. 
\section{Considerations for Cavities}
 We now step back from the details of the comparisons described above and reassess the overarching question for the prominence-cavity system: what do the observations of prominence horns imply about the thermodynamic state of the cavity? In one school of thought, the cavity density is depleted because it feeds mass into the prominence \citep[e.g.,~][]{saitotand_73, fort_74, berger_12}.
This hypothesis is based on two distinct underlying assumptions: a condensation process occurs within the cavity, and that process only converts mass from the coronal cavity into prominence plasma.
The results of Section 3 suggest that horns are formed by cooling plasma, and horns appear to extend from the cavity into the prominence.
Can this cooling process, when occurring over many loops, produce a large-scale density-depleted cavity?\indent

To directly answer this question, it is necessary to compare the plasma properties of the cavity with those of the predicted by our loop simulations.
However, it is difficult to find common ground for a quantitative comparison, due to key model assumptions.
First, the loop geometry and the heating parameters applied in our model are suitable for studying the cooling and equilibration of a gravitationally stable condensation, whereas observed prominences and cavities typically are dynamic \citep{liu_12}.
Second, the condensation in the present simulations is produced by localized steady heating, whereas the actual temporal and spatial structure of coronal heating is a lively topic of debate \citep[e.g.,~][]{klimchuk_06}. 
The magnitude of the applied footpoint heating is within the range of canonical estimates \citep{withbroe_77} but this provides very loose constraints. 
In order to ultimately understand the time-averaged loop properties, we need more concrete information on the coronal heating mechanism itself and the lifecycle of prominence threads.\indent
\begin{table}
\begin{center}
\label{tab:1}
\begin{tabular}{r|ccc}
 & Run A & Run B & Horns\\
 \hline
log  $E_1$& -2.8 & -3.7& -\\
 T$_{MAX}$ [MK]& 3.4 & 1.7& -\\
 t$_{193}$ [hr]& -0.4& -2.9 & -0.5$\pm0.4$\\
 $\Delta t_{193}$[hr]&3.8&2.2&3$\pm0.4$\\
 $\Delta I_{193}$&1.50&1.01 & $<$1.2\\
 t$_{171}$[hr]&0.&-0.3&-0.5$\pm$0.3\\
  $\Delta t_{171}$[hr]&2.1&7.2&3$\pm0.3$\\
 $\Delta I_{171}$&11.0&1.8&1.2-1.8\\
  \end{tabular}
  \end{center}
 \caption{Comparison of model values with observations of horns. }{$T_{MAX}$ is the maximum temperature reached during the simulation. $t_{193}$ is the time of peak emission in the 193\AA~bandpass, $\Delta t_{193} $ is the duration of the brightening, $\Delta I_{193}$ is the maximum intensity of the brightening relative to background. All quantities subscripted ``171" label congruent statistics but for 171\AA~emission.}
 \end{table}

Our simulations of TNE predict that condensation formation reduces the density in the coronal segment of the cooling loop, but only after the density has been increased through chromospheric evaporation. 
After a condensation has formed, the coronal portions of prominence-containing loops exhibit lower densities (70\% at loop apex) and lower temperatures (50\% at loop apex) than the \cite{rosner78a} scaling laws would predict for a half-loop of the same energy input. 
The TNE process can occur regardless of loop geometry, as long as the length-to-heating scale exceeds $\sim10$ \citep{karpen_01}. 
However, coronal rain is not prevalent in streamers, so  the cavity-streamer density difference cannot be explained simply by invoking thermal nonequilibrium. 
One possibility is that different heating mechanisms operate in streamers and cavities, but thus far there is no compelling observational evidence for this suggestion.
A change in magnetic structure also could differentiate between the cavity and the streamer.
According to the \cite{rosner78a} scaling laws, there is a weak direct relationship between loop length and density.
These scaling laws do not apply to loops with the localized footpoint heating of the TNE model, however, or to dynamically evolving loops. 
To resolve this important issue, we need to determine whether the heating in the cavity differs significantly from that in the streamer, or whether geometrical effects can explain the difference.\indent

Better understanding of projection effects and filling factors also are important. 
Cavities can span hundreds of megameters in the axial direction; given the dispersed distribution of high latitude photospheric flux, a single large-scale cavity might encompass several distinct magnetic flux systems.
In the synthetic emission shown in Section 3, the temporal and spatial averaged measurements were used to produce an emission background, but a more realistic approach would consider the filling factor of the condensing versus non-condensing field lines. 
Our prior multithread studies of prominences indicate that filling factors of order 0.001 yield reasonable agreement with estimated masses of observed prominences \citep{luna_12}.\indent

One avenue of study that is worth pursuing is to probe for the presence of a relatively-hot ($> 2$ MK) component in the cavity. In most of the condensing loops studied here, the maximum loop temperature achieved prior to the cooling phase exceeds 2 MK (Run A reaches 3.4 MK).
This phase of the evolution may occur at a significantly lower density (depending on the equilibrium heating) than the cooling or post-condensation phase (see Figure~\ref{fig:evo}), thus producing a weak EUV emission signature.
The hot bandpasses available to AIA, specifically 335\AA~and 131\AA, exhibit low signal in the quiet Sun and the cavity. A long-exposure spectroscopic study using Hinode/EIS would allow us to set limits on the presence and variability of $>2$MK plasma surrounding the prominence.\indent

This study suggests that horns are the manifestation of cooling plasma adding mass to the prominence. Based on the line-of-sight projection, these features appear inside the cavity. The TNE model predicts that a density depletion will occur in the corona that coincides with the formation of a condensation.
While it is possible this depletion is responsible for the existence of the cavity, more evidence is needed to constrain the physical properties of the cavity and the streamer.

\section{Discussion}
This analysis has combined two distinct but inherently related areas of prominence research: dynamic modeling and time-dependent EUV observations. This type of comparison is an essential step toward understanding the prominence-cavity system. We have attempted to present the predicted and observed datasets in a congruent manner, and have found that our theoretical models are a close but inexact fit to the observations. \indent

We have presented evidence that the dynamic emission observed surrounding prominences and projecting inside the cavity has several characteristics that are compatible with the TNE model for a catastrophically cooling loop.
The TNE provides relevant predictions on the thermodynamic state of the cavity, but a more thorough comparison requires a more detailed understanding of the heating mechanism and magnetic structure within and surrounding the prominence. The evidence at hand suggests that horns are related to the formation of prominence mass, and that they connect the cavity and the prominence.\indent

One area of concern in our comparison of the TNE and the horn observations comes from the timing correlations between the various emission structures.
Our simulations never exhibit 171\AA\ or 193\AA\ emission that peaks {\it after} the formation of the condensation (such as Horn 3, Figure~\ref{fig:horn}).
One process that might produce this situation is a slowly cooling loop which undergoes a rapid mass increase coincidental with the formation of the condensation. Another possible scenario is that the line of sight intersects multiple loops in different stages of TNE, complicating the interpretation of the emission profiles from the horns, prominence, and cavity. Further study of our multithread prominence simulations could offer a more realistic approach to this 3D problem \citep[e.g.,~][]{luna_12}.
\indent

Many avenues remain open to be explored in our understanding of the prominence environment. The next step forward in our theoretical understanding is the self-consistent modeling of the magnetic structure and energetic evolution simultaneously, but this is a computationally challenging problem due to the large range of spatial and temporal scales that must be modeled.
Moreover, while our 1D loop simulations have expanded our understanding of the plasma dynamics and radiative signatures, the necessity of supplying such models with an empirical heating function will always limit the scope of their results. 
\cite{choe_92} offered a preliminary attempt at the full magnetic-plasma problem, namely that field line extension can siphon mass into the corona.
\cite{low_12} suggests that cross field plasma transport may be important, highlighting another process that 1D models cannot capture.
It remains to be seen how the current generation of MHD models will elucidate the problem.
The fundamental process described by TNE is that mass-loading the corona by evaporation of chromospheric plasma leads to cooling.
Why, then, does this process occur in commonly in filament channels but infrequently in the surrounding streamer? Forthcoming observational studies of filament channel and prominence evolution will shed light on the connection between magnetic structure and plasma dynamics, thus also illuminating the underlying physics of coronal heating and prominence formation.
\\
\\
{\it This work was partially funded by the Max-Planck/Princeton Center for Plasma Physics.
The National Center for Atmospheric Research is funded by the National Science Foundation.
ML gratefully acknowledge partial financial support by the Spanish Ministry of Economy through projects AYA2011-24808 and CSD2007-00050. This work contributes to the deliverables identified in FP7 European Research Council grant agreement 277829, Ó Magnetic connectivity through the Solar Partially Ionized AtmosphereÓ, whose PI is E. Khomenko.
}

\bibliography{mybibliography}

\end{document}